\documentclass[useAMS, usenatbib, fleqn]{mn2e}
\usepackage{graphicx}
\usepackage{epstopdf}
\usepackage{amsmath}
\usepackage{multicol}
\usepackage{amssymb}

\def\apj{APJ}

\def\apjl{APJ, Part 2 - Letters}
\def\aap{Astronomy and Astrophysics}

\def\mnras{MNRAS}
\def\nat{Nature}

\def\prd{Physical Review D}
\def\physrep{Phys. Rep.}

\title{Roche Accretion of stars close to massive black holes}

\author[Lixin (Jane) Dai, \& Roger Blandford]{Lixin Dai$^{1}$\thanks{E-mail:
cosimo@stanford.edu (LD)}, and Roger Blandford$^{2}$\thanks{E-mail: rdb3@stanford.edu (RB)} \\
$^{1,2}$Kavli Institute for Particle Astrophysics and Cosmology, Stanford University, Menlo Park, CA 94025, USA\\}

\begin{document}

\pagerange{\pageref{firstpage}--\pageref{lastpage}} \pubyear{2009}

\label{firstpage}

\maketitle

\begin{abstract}

In this paper we consider Roche accretion in an Extreme Mass-Ratio Inspiral (EMRI) binary system formed by a star orbiting a massive black hole. The ultimate goal is to detect the mass and spin of the black hole and provide a test of general relativity in the strong-field regime from the resultant quasi-periodic signals. Before accretion starts, the stellar orbit is presumed to be circular and equatorial, and shrinks due to gravitational radiation. New fitting formulae are presented for the inspiral time and the radiation-reaction torque in the relativistic regime. If the inspiralling star fills its Roche lobe outside the Innermost Stable Circular Orbit (ISCO) of the hole, gas will flow through the inner Lagrange point (L1) to the hole. We give new relativistic interpolation formulae for the volume enclosed by the Roche lobe. If this mass-transfer happens on a time scale faster than the thermal time scale but slower than the dynamical time scale, the star will evolve adiabatically, and, in most cases, will recede from the hole while filling its Roche lobe. We calculate how the stellar orbital period and mass-transfer rate will change through the ``Roche evolution" for various types of stars in the relativistic regime. We envisage that the mass stream eventually hits the accretion disc, where it forms a hot spot orbiting the hole and may ultimately modulate the luminosity with the stellar orbital frequency. The observability of such a modulation is discussed along with a possible interpretation of an intermittent 1 hour  period in the X-ray emission of RE J1034+396.
\end{abstract}


\section{Introduction}
   The measurement of black hole parameters - mass and spin - has been an important astrophysics research topics over the past decade. Such measurements can lead to tests of general relativity in the strong-field regime, and reveal information on the formation and evolution of galaxies. There has been great progress on this topic using various approaches. The mass of a black hole is more easily measured since it is a Newtonian parameter. However, the spin, a relativistic parameter, is much harder to be measure. One method is to use radiation from the accretion disc: either by fitting the shifts and shapes of the iron lines (e.g., \citet{Reynolds03}), or by using a multi-temperature blackbody model \citep{Yuan04, Li09, Shcherbakov10, Gou11}. The problem with the iron line model is that it is not certain that the Fe lines indeed come from near the Innermost Stable Circular Orbit (ISCO). The continuum method can be employed to obtain the spin of stellar mass black holes effectively. However, for Super-Massive Black Holes (SMBHs), the radiation from the inner disc region is in the UV range and can be absorbed. Jets are commonly associated with black hole spins (e.g., \citet{Blandford77, Tchekhovskoy11}). However these models are more complicated. Here we propose another way of detecting SMBH spins using tidal effects induced by star-SMBH close passages. 

   There are two extreme cases of such tidal effects. The first case is a tidal disruption. This can happen when a star passes close by a black hole in an unbound orbit or a very eccentric orbit. The stellar volume exceeds the Roche volume at periastron and the star is torn apart and eventually swallowed by the hole \citep{Rees88, Stone11}. One possible instance of a tidal disruption is the recent Sw J16444$+$57 observation\citep{Bloom11, Zauderer11}. These events are very luminous but hard to model. In the other case, a star orbiting the hole in an equatorial circular orbit spirals inward under the action of gravitational radiation. When the star fills it Roche Lobe, mass is stripped stably from the star on a time scale of $\sim 10,000 - 100,000 $ years. This steady mass-transfer is not as luminous as tidal disruption, but lasts for much longer. Also the second case is easier to model with high precision for the mass and spin of the hole. In this paper we will provide techniques for doing so.
   
   Mass-transfer in binary systems has been well studied for a variety of stellar components and mass ratios. This investigation is mostly focused on binary systems composed of a stellar-mass compact object and a companion star, e.g. \citet{Webbink85}.  In particular, \citet{Hjellming87} and  \citet{Soberman97} explored the stability of adiabatic mass-transfers based on polytropic stellar models, and showed it could be dynamically unstable. However, these studies are Newtonian calculations, since the companion star reaches the Roche limit far away from the ISCO of the stellar-mass compact object. In Extreme Mass-Ratio Inspiral (EMRI) systems, stars can get close to the ISCO before being tidally disrupted. Therefore, the signals from such binary systems can in principle be used to constrain the mass and spin of SMBHs, and to test general relativity in the strong-field regime. In this paper, we present an accurate general relativistic framework for studying such mass-transfer.
   
   We shall be especially concerned when the mass-transfer is ``adiabatic" - in other words it happens on a time scale short compared with the thermal (Kevin-Helmholtz) time for the star yet long compared with the dynamical time. This problem has previously been considered by \citet{Hameury94} in connection with the (erroneous) identification of periodicity in NGC6814 \citep{Madejski93}. This paper builds upon the \citet{Hameury94} discussion and presents a more comprehensive treatment of stars and planets. In particular, we will show for various stellar types how the stellar Roche volume and evolution change with the spin parameter of the hole and also the rotational angular velocity of the star in the relativistic regime. We will also conclude that such accretion events are stable over a dynamical time scale, in contrast to the suggestions of \citet{Hameury94}. The differences in results are mainly caused by more accurate calculation of the relativistic Roche potential (and the resultant Roche volume), as well as more accurate gravitational radiation power calculation.

    In an EMRI system, the central black hole is so massive that its mass is effectively constant. The  stellar orbit, on the other hand, shrinks due to the loss of energy and angular momentum through gravitational radiation and other torques. This radiation is a prime target for space-borne gravitational radiation observatories intended to test the general theory of relativity in the strong field regime. An EMRI system can also emit electro-magnetic radiation (principally X-rays and optical), and its observation can also provide the basis of a good test of general relativity, if measurements are carried out accurately. 
        
    We restrict our discussion of EMRI systems to stars in circular, equatorial orbits. If and when such a star overflows its Roche lobe close to the ISCO, gas will flow through the inner Lagrange point (L1), and the stellar orbit will usually expand - an Extreme Mass Ratio Outspiral or EMRO! We shall assume initially that the stellar evolution is adiabatic, and discuss later the limitations of this assumption. Under these conditions, the entropy variation with mass within the star is frozen and the evolution is as described in our previous paper. In the Newtonian regime it is the sign of $({\partial \bar{\rho_\star}} / {\partial m_\star})_s$, where $m_\star$ is the total remaining mass of the star, and $\bar{\rho}$ is the mean stellar density, that determines whether or not an EMRI becomes an EMRO.     
     
    The gas flowing through L1 will generally have enough angular momentum to form a torus orbiting the hole. A hot spot will be formed where the stream hits the disc. In addition, it is quite possible that one or two spiral arm shock waves will be created in the disc. The hot spot and shocks will orbit with the orbital frequency of the star and an inclined observer should see a modulation of the radiation from the disc, most likely in the X-ray range. Clearly the modulation will be the greatest when the observer is in the equatorial plane, observing maximal Doppler boosting.
    
    One observation that may possibly be explained by this model is the quasi-periodic XMM-Newton X-ray signal from a nearby active Seyfert I galaxy RE J1034+396 observed by \citet{Gierlinski08}. As shown in Fig. 1 in their paper, the rest-frame X-ray emission has a periodicity of $\sim$ 1 hr. The X-ray flux has a modulation of $\sim 0.1$ associated with the nucleus of the galaxy, and this modulation could be well fitted by a cosine function. We will show in this paper that a dwarf star or brown dwarf or red giant can produce emissions with such a periodicity filling a $\sim 10^6-10^7 M_\odot$ SMBH under Roche accretion, though the mass accretion rate under adiabatic condition and pure gravitational radiation torque is too low to produce the observed X-ray flux.

     In section 2, we will illustrate the mechanism by computing the evolution in the Newtonian limit. In particular, we will show how the star spirals in, when it reaches the Roche limit, and we will calculate how the stellar orbital period and mass-transfer rate evolve taking a dwarf star as representative. We will also  discuss the final fate of stars after they have undergone Roche evolution. In section 3,  these calculations will be repeated in the relativistic regime, and be compared with \citet{Hameury94}'s. In section 4, we will use the data from RE J1034+396 to fit our model, and to make predictions of the future signals. In particular, we will discuss how spins can be constrained from signals. Discussion and future research directions will be laid out in the last section. In our forthcoming paper in preparation, we will use these new results to construct new models of EMRI/EMRO evolution.

\section{Newtonian Treatment}

A Newtonian treatment brings out the salient features of the problem, e.g., \citet{Hameury94}. Consider a fully convective lower main sequence star with a mass of $m_\star  \equiv  m M_\odot $ as a representative, and assume that it orbits a hole with a mass $M \equiv  10^6 M_6 M_\odot$, in a circular, equatorial orbit with radius $r$. (The stellar orbit circularizes quickly either through gas dynamical drag or gravitational radiation as shown in \citet{Dai11a}. We will take $G = c =1$ henceforth.

From Kepler's third law, the orbital radius (measured in units of the gravitational radius $ R_g (M_6) = 1.47 \times 10^{11} M_6 \ \rm{cm}$) is:

\begin{equation}
  \label{PeriodNewtonian} 
  \tilde{r}= \frac{r}{R_g} = 23.9 \ P^{\frac{2}{3}} M_6^{-\frac{2}{3}},
\end{equation}
where the orbital period $P$ is measured in hours.

The orbital angular momentum:
\begin{equation}
  \label{LNewtonian} 
  L=  m_\star \sqrt{M r} = 4.9  \ m  M_6^{\frac{2}{3}} P^{\frac{1}{3}}
\end{equation}
measured in units of $M_\odot \ c \ R_g (M_6=1) = 8.8 \times 10^{54} \rm{g \ cm^2 / s}$.

Likewise, the orbital energy, in units of $M_\odot c^2 = 1.79 \times 10^{54}$ erg, is
\begin{equation}
  \label{ENewtonian} 
  E = -0.5 M m_\star r^{-1} = -0.021 \times m M_6^{\frac{2}{3}}P^{-\frac{2}{3}}.
\end{equation}

\subsection{Pre-Roche evolution}
We assume that the star follows a circular orbit which shrinks slowly due to gravitational radiation. In the Newtonian limit, the torque is
\begin{equation}
  \label{TorqueNewtonian} 
  \tau = \frac{32}{5} M^{\frac{4}{3}} {m_\star}^2 (\frac{2\pi}{P})^{\frac{7}{3}} = 1.7\times10^{44}m^2 {M_6}^{\frac{4}{3}} P^{-\frac{7}{3}} \rm{g \ cm^2 s^{-2}}.
\end{equation}

The time it takes the star starting from radius $r$ to plunge into the hole is (assuming this gravitational radiation torque, and $r \gg r_{\rm ISCO}$, which is the radius of the ISCO, see below):
\begin{equation}
 \label{InspiralTimeNewtonian} 
t_{\rm{inspiral}} = \frac{5}{256} \frac{r^4}{M^2 m_\star} \sim 987 \ m^{-1} M_6^{-\frac{2}{3}} P^{\frac{8}{3}} \  \rm{yr}.
\end{equation}
(e.g.  \cite{MTW}).

Typically if the orbit is at least mildly relativistic, the inspiral time is less than a Kelvin-Helmholtz time.

\subsection{Roche limit}
In order to obtain the Roche volume of the star, it is convenient to transform from an inertial frame to a frame of reference rotating with the binary system.

The Roche potential $\Phi (\vec{r})$ for an arbitrary point with position vector $\vec{r}$ in the rotating frame is given by:

\begin{equation}
 \label{RochePotentialNewtonian} 
 \Phi (\vec{r})_N= - \frac{m_\star}{| \vec{r}-\vec{r}_\star |} - \frac{M}{| \vec{r}-\vec{r}_{\rm BH} |} - \frac{1}{2} |\vec{\omega} \times \vec{r}|^2
\end{equation}
\cite{Frank}, where $\vec{r}_\star$ and $\vec{r}_{\rm BH}$ are the position vectors of the centers of the star and the hole, and $\omega$ is the rotational angular velocity of the star in the inertial frame. $\omega$ satisfies $0 \leq \omega = k_{\omega} \Omega \lesssim \Omega$, where $\Omega$ is the stellar orbiting angular velocity. 

The first term in this equation in the Roche potential is the gravitational potential of the star itself, making the approximation that the stellar mass is centrally concentrated and only retaining the monopolar term. The second term is the gravitational potential term of the hole. The third term is the potential in the rotating frame due to the centrifugal force.

We erect a locally orthonormal coordinate system $\{x, y, z \}$ centered at the star and directed along the $\{ r, \theta, \phi \} $ coordinate axes. If we expand the gravitational potential near the star,  and note that the linear term is canceled by a linear term in the expansion of the centrifugal potential, the Newtonian tidal tensor contributes quadratic terms $-\frac{1}{2} (\phi_{, xx} x^2 +\phi_{, yy} y^2 +\phi_{, zz} z^2)$, where $\phi_{, xx} = -2 \phi_{, yy}= -2 \phi_{, zz} = 2 M / r^3$. (Note that the contribution to the Laplacian vanishes as it should.) The expansion of the centrifugal term also contributes a quadratic form $-\frac{1}{2} \omega^2 (x^2 + z^2)$, or $-\frac{1}{2} {k_\omega}^2 (x^2 + z^2) M/r^3$.

A co-rotating star will expand to fill an equipotential surface. However, it is rather unlikely that an inspiralling star will be able to maintain corotation. If it rotates with angular frequency $\omega < \Omega$, the quadratic centrifugal term is approximately $-\frac{1}{2} \omega^2 (x^2+z^2)$. which includes the effects of Coriolis force. (There may be some internal circulation which will vitiate adiabaticity, but we do not consider this possibility further in this paper.)

The gradient of the Roche potential vanishes at the first and second Lagrange point.  Using this we can calculate for two extreme cases that the first and second Lagrange points (L1 and L2) are located at
 \begin{equation}
  \label{L1} 
  x_R = \pm 0.794 \  \mathcal{X}(k_\omega)  \left (\frac{m_\star}{M} \right )^{\frac{1}{3}} r
 \end{equation}
from the star, where $\mathcal{X}(k_\omega)$ decreases from 1 to 0.87 as $k_\omega$ goes from 0 to 1.

The transverse extents of the equipotential Roche surface are:
\begin{equation}
  \label{y_R} 
  y_R = \pm 0.497 \ \mathcal{Y}(k_\omega) \left (\frac{m_\star}{M} \right )^{\frac{1}{3}} r 
\end{equation}
and
\begin{equation}
  \label{z_R} 
  z_R = \pm 0.497 \ \mathcal{Z}(k_\omega) \left (\frac{m_\star}{M} \right )^{\frac{1}{3}} r 
\end{equation}
where $\mathcal{Y}(k_\omega)$ goes from 1 to 0.89 and $\mathcal{Z}(k_\omega)$ goes from 1 to 0.93 as $k_\omega$ goes from 0 to 1.

Furthermore, the volume contained by the Roche surface is numerically calculated  to be:
\begin{equation}
  \label{VN} 
  v_N \simeq  0.683  \times \left ( \frac{k_\omega}{2.78} +1 \right )^{-1} \frac{m_\star}{M} r^3
\end{equation}

Combining this with Eq. [\ref{PeriodNewtonian}], we obtain the relationship between the orbital period and the mean density of the star when the star reaches the Roche limit:

\begin{equation}
  \label{RochePeriodNewtonian} 
  P_0 \simeq 8.18  \left ( \frac{k_\omega}{2.78} +1 \right )^{\frac{1}{2}} {\bar{\rho}}^{-1/2} 
\end{equation}
independent of the mass of the hole,where $P_0$ is the stellar orbital period in hours, and $\bar{\rho}$ the mean density of the star in cgs unit. For a dwarf star with mass $\sim 0.3 M_\odot$ , the Roche condition is satisfied when the orbital period decreases to a value of $ 1.96 - 2.27$ hr depending on $k_\omega$. If we suppose that the star plunges when its orbital radius reaches $6 R_g$, the maximum black hole mass for there to be mass-transfer from a $\sim 0.3 M_\odot$ dwarf star is $1.6 - 1.9  \times 10^7 M_\odot$ ($ k_\omega = 0 -1$).

\subsection{Roche evolution}

To set the stage for the relativistic treatment below, let us consider the classical Roche model for a dwarf star that is co-rotating. Its mean density decreases monotonically as mass is tidally stripped, leading to expanding orbits. Assuming that the star has initial mass $m_{\star 0} \sim 0.3 M_\odot$, we found that

\begin{equation}
  \label{mrho} 
 \frac{ \bar{\rho}_\star} {\bar{\rho}_{\star 0}} \simeq  \left( \frac{m_\star}{m_{\star 0} } \right)^{2.2},
\end{equation}
(\citet{Dai11a}), where $\bar{\rho}_{\star 0}$ is the initial mean density of the star.

Combining Eq. [\ref{RochePeriodNewtonian}] and [\ref{mrho}], the orbital period $P$ of the dwarf star, in terms of the initial period $P_0$, varies with its remaining mass as:
\begin{equation}
  \label{Pm} 
 \frac {P} {P_0} \simeq  \left( \frac{m_\star}{m_{\star 0} } \right)^{-1.1}.
\end{equation}

We assume that the orbit of the star will change in a way that the angular momentum of the system is conserved so that
\begin{equation}
  \label{Lconservation} 
 \tau + \frac{dL}{dt}=0,
\end{equation}
where we continue to assume that gravitational radiation provides the only torque as in Eq. [\ref{TorqueNewtonian}].

Using Eq. [\ref{LNewtonian}], [\ref{TorqueNewtonian}], [\ref{Pm}], and [\ref{Lconservation}], we see the time elapsed since the onset of mass-transfer goes as:
\begin{equation}
  \label{timeR} 
 t_R  \sim 4.1 \times 10^3  M_6^{-\frac{2}{3}} {P_0}^{2.7} \left( \left( \frac {P} {P_0} \right)^{3.6}- 1 \right) \ {\rm yr} .
\end{equation}

The associated rate of mass transfer of Roche accretion, also obtained assuming conservation of angular momentum, is:
\begin{eqnarray}
  \label{mdotN} 
 \dot{m}_\star  & \sim & - 1.23\times10^{22}  M_6^{\frac{2}{3}} m^2 P^{-\frac{8}{3}} \nonumber \\
                      & \sim & -3.62 \times10^{23} {P_0}^{-2.7} M_6^{0.7} m^{4.9} \rm{g \ s^{-1}}.
\end{eqnarray}

\begin{figure}
 \centering
 \includegraphics[width=3in]{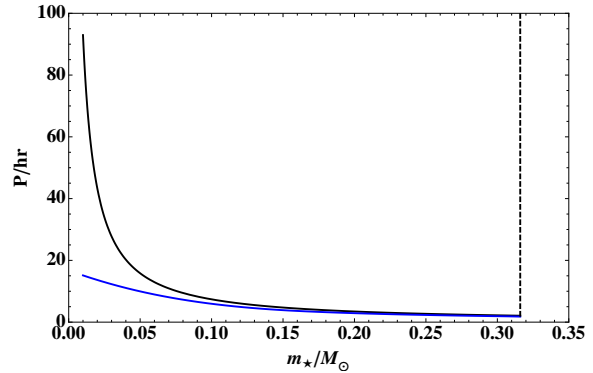}
 \caption{
This figure shows the period-mass relationship of a non-rotating lower main-sequence star with initial mass $\sim 0.3 M_\odot$ accreting onto a Schwarzschild SMBH with a mass of $10^7 M_\odot$. The vertical dashed line represents the evolution prior to the mass-transfer, when the mass does not change, but the orbit shrinks. The black curve describes the subsequent Roche evolution in the Newtonian limit, when the stellar orbit expands. The blue curve shows how stellar orbital period changes with mass in the relativistic limit.} 
 \label{PMlm}
\end{figure}

\begin{figure}
 \centering
 \includegraphics[width=3in]{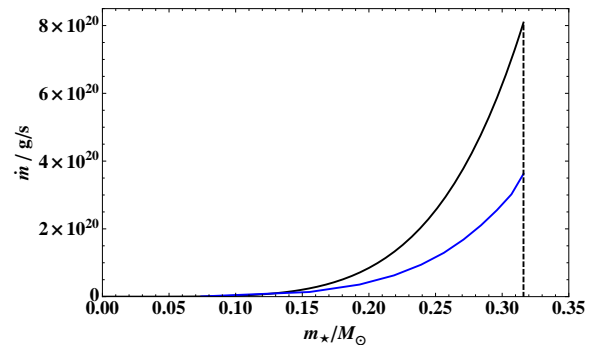}
 \caption{
This figure shows how fast a non-rotating lower main-sequence star with initial mass $\sim 0.3 M_\odot$ accretes onto a Schwarzschild SMBH with a mass $10^7 M_\odot$ through the Roche evolution. The black and blue curves are Newtonian and relativistic calculations correspondingly. }
 \label{Mdotlm}
\end{figure}

We show in Figs. \ref{PMlm} and \ref{Mdotlm} the Newtonian evolution of this low-mass star in orbit around a hole with mass $10^7 M_\odot$. We see that the stellar orbital period can increase to around five times the Roche period as the star is stripped. However, the majority of the Roche evolution happens on a time scale $\sim 10^5$ years, and the stellar mass-transfer rate deceases rapidly with time.

As discussed by \citet{Dai11a}, the angular momentum of the binary system as in Eq. [\ref{LNewtonian}] follows $L \sim m_\star P^{\frac{1}{3}} \sim m_\star {\bar{\rho}_\star}^{-\frac{1}{6}} $. If the mass accretion is stable on dynamic time scales, it mean $L$ should decrease as mass is accreted onto the hole. If we define a parameter $\eta = d (\ln \bar{\rho}_\star) / d (\ln m_\star )$, then $\eta <6 $ is the criteria for Newtonian dynamical stability.  

A more intuitive method of calculating the dynamical stability, as \citet{Hameury94} suggested, is to ensure that the star will not exceed the Roche surface under the assumption of no angular momentum loss through gravitational radiation. In other words, $d v_\star / d m_\star > d v_{\rm{N}} / d m_\star$, where $v_\star = m_\star / \bar{\rho}_\star$ is the stellar volume, and $v_{\rm{N}}$ is the Roche volume in Eq. [\ref{VN}]. This leads to the same stability criteria $\eta<6$.

Continuing to consider a $\sim 0.3M_\odot$ dwarf star, we have $\eta = 2.2$ (Eq. [\ref{mrho}]), ensuring that the star recedes from the hole stably under Roche accretion.

For many purposes this Newtonian treatment applied to a broader range of stars will be adequate. However, as we want to exhibit relativistic effects, we will turn to a relativistic treatment, again building upon the discussion in \citet{Hameury94}.

\subsection{ The final fate of stars under Roche evolution}

After mass-transfer begins, most stars will recede from the black hole so that they will continue to fill their Roche lobes as their density falls. The star loses mass less and less rapidly. Eventually the star will cool and its structure will evolve. The motion of the star under the action of gravitational radiation will then cause the orbit to shrink. Additional excursions are possible depending upon the structure of the star. The star might cool down adiabatically or on a thermal time scale. Therefore, it might or might not fill it Roche Lobe on its return. The equilibrium properties of these stars will be even more complicated than the ones we calculated in this paper. Ultimately the star crosses the ISCO and plunges into the hole. When more than one star is in a ``parking orbit'' there is the possibility of resonant dynamical interaction analogous to that which occurs in planetary systems. In addition interaction with the disc can alter the evolution.

We illustrate the scenario in the following Fig. \ref{Parking}  Around the hole, there will be stars migrating inward and mostly outward under Roche evolution,  stars that have gone at least one Roche evolution ``parked" in a belt or cloud (if the stars came in on inclined orbits) waiting to move in again, new stars coming in under gravitational radiation still burning, and stars coming in from the ``parking cloud" and not burning.   

 \begin{figure}
 \centering
 \includegraphics[width=3in]{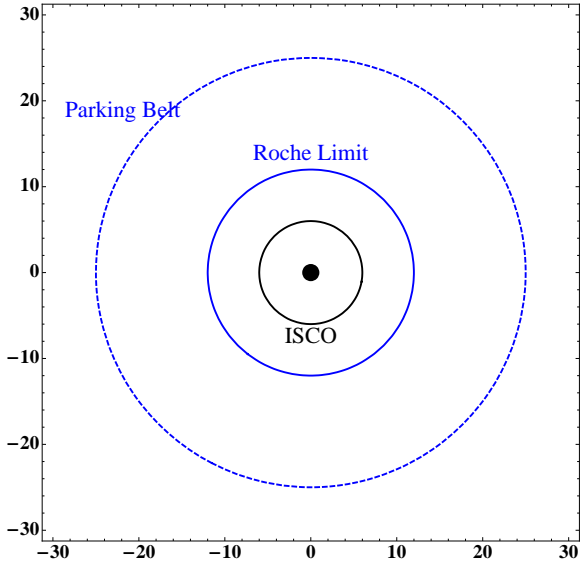}
 \caption{For a $\sim M_\odot$ star orbiting a Schwarzschild SMBH with mass $10^7 M_\odot$ as discussed, the star reaches the Roche limit when the orbital radius is $\sim 12 R_g$, and then spiral out. When the evolution time scale is comparable with Kevin-Helmholtz time scale, the star stops receding in the ``parking belt'', which is located at around $\sim 25 R_g$ from the hole. Here we calculated the Kevin-Helmholtz time scale of the star using the initial stellar internal energy and luminosity. }
 \label{Parking}
\end{figure}

\section{Relativistic treatment}

\subsection{Gravitational radiation emission}

A star (with a mass $m_\star$) orbiting a massive hole (with a mass $M$ and a spin parameter $a$) will lose angular momentum and energy due to gravitational radiation. The orbit is circularized quickly, as we have shown in \citet{Dai11a}. However the rate of alignment of the spin and orbital angular momentum is slower than the inspiral rate (e.g., \citet{Hughes01}). For the moment we restrict attention to circular equatorial orbits. 

We define a dimensionless orbital angular velocity $\tilde{\Omega} = M\Omega$, where $\Omega$ is the regular orbital angular velocity. The relativistic Kepler's law is:
\begin{equation}
  \label{AngularVelocity} 
   \tilde{\Omega} = \frac{1}{\tilde{r}^{\frac{3}{2}} \pm a},
\end{equation}
where $\pm$ refers to prograde and retrograde orbits.

The star plunges into the hole when the orbit shrinks to the ISCO. The ISCO of a black hole of mass $M$ and spin parameter $a$ has a radius $r_{\rm{ISCO}}$ that satisfies the equation as given in \citet{Bardeen}:

\begin{equation}
\left(\frac{r_{\rm{ISCO}}}{M}\right)^2-6 \frac{r_{\rm{ISCO}}}{M}+8 a \sqrt{\frac{r_{\rm{ISCO}}}{M}}-3a^2=0.
\end{equation}
This radius equals $R_g$, $6 R_g$, and $9 R_g$ for $a = 1, \ 0,$ and $-1$ respectively. The upper limit for $a$ has been argued to be $\sim 0.998$  \citep{Thorne:74} for which $r_{\rm{ISCO}} \sim 1.237 R_g$.

\begin{figure}
 \centering
 \includegraphics[width=3in]{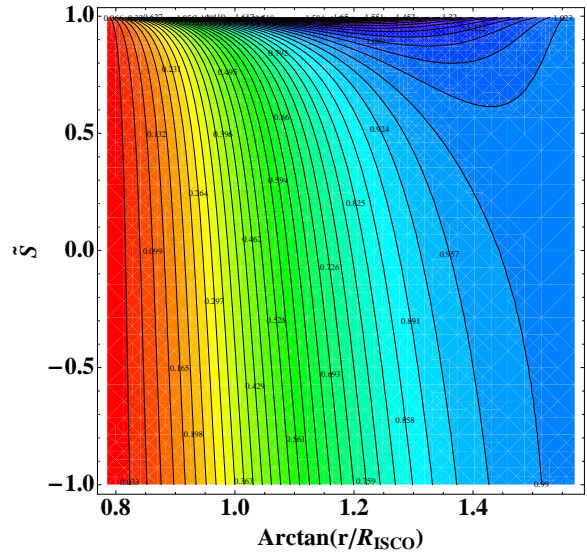}
 \caption{
The relativistic correction term $\mathcal{T}$ to the plunge time as a function of the distance to the hole and the spin parameter of the hole. We see that as the radius goes to infinity, $\mathcal{T}$ becomes 1 no matter what the spin is.  Also $\mathcal{T} = 0$ when $r= R_{\rm ISCO}$.} 
 \label{contourTau}
\end{figure}

Let us now consider how to modify the treatment of gravitational radiation. \citet{Finn} computed a relativistic inspiral time scale for a star at an orbital radius $r$:

\begin{equation}
 \label{InspiralTime} 
   t_{\rm{inspiral}} = \frac{5}{256} \frac{M^2}{m_\star \tilde{\Omega}^{\frac{8}{3}}} \mathcal{T}
\end{equation}
in the Boyer-Lindquist coordinate, where $\mathcal{T}$ is the relativistic correction term. They presented their results of $\mathcal{T}$ in tabular form. We have used these results to give a convenient interpolation formula for $\mathcal{T}$:

\begin{eqnarray}
\label{InspiralTimeCorrection} 
\mathcal{T}(x,a) &=& \frac{(x - 1)^2 + b \ (x - 1)}{x ^2 + d \ x  + f} ,  \\
b &= & 1.06 \times (1-a)^{-0.11}-0.66,   \nonumber \\
d &=&   0.57 \times (1-a)^{0.23} - 2.22,  \nonumber \\
f &=&   [-1.08 \times (1-a)^{1.08} - 3.18 \times (1-a)^{-0.018}]  \nonumber \\
           &&  \times e^{-0.23\times(1-a)} +4.91, \  \nonumber
\end{eqnarray}
where $x = r/r_{\rm ISCO}$. Clearly $T(x=1, a) =0$, and $T(x, a) \rightarrow 1$ as $x \rightarrow + \infty$ (since at infinity the relativistic effects are negligible). This formula holds true with a maximum error of $\sim 2\%$ for $x\ge2$. A contour plot of $\mathcal{T}(x,a)$ is shown in Fig. \ref{contourTau}.

\begin{figure}
 \centering
 \includegraphics[width=3in]{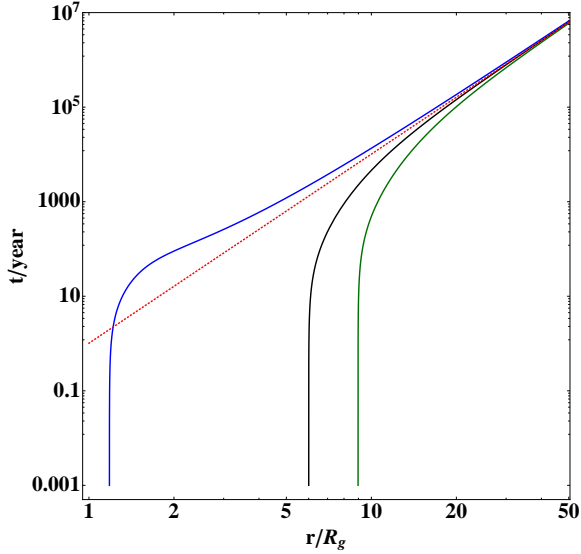}
 \caption{
The total inspiral time in years as a function of $r/R_g$  for a $0.3 M_\odot$ dwarf star orbiting a $10^7M_\odot$ hole. This plot is in the log-log scale. The three colors indicate three different spin parameters of the hole:  -0.99 (green), 0 (black), and 0.999 (blue). The three solid curves are accurate inspiral times obtained using Eq. [\ref{InspiralTime}]. The red dotted curve is the Newtonian inspiral time  using Eq. [\ref{InspiralTimeNewtonian}]. All other curves approach the Newtonian curve when the star is far away from the hole.  } 
 \label{InspiralTimePlot}
\end{figure}

The results for the evolution of the $0.3 M_\odot$ dwarf star are shown in Fig. \ref{InspiralTimePlot}. It is clear that the standard "Landau-Lifschitz" form equation  is adequate for $r > 100 R_g$.

The gravitational radiation reaction torque is readily computed from:

\begin{equation}
 \label{torquer} 
   \tau (r, a)_{\rm{gr}}= - \left (\frac{\partial L}{ \partial r} \right) _a  /  \left( \frac{\partial {t_{\rm{plunge}}} }{\partial r} \right) _a,
\end{equation}
where the orbital angular momentum 
\begin{equation}
 \label{Lr} 
L= m_\star M \frac { ( 1 - 2 \tilde{r}^{-1} + a*\tilde{r}^{- \frac{3}{2}} ) \  r^{\frac{1}{2}}} {(1 - 3 \tilde{r}^{-1} + 2 a*\tilde{r}^{- \frac{3}{2}})^{\frac{1}{2}} }
\end{equation}
(e.g., \citet{Bardeen}).

\subsection{Roche limit}

We use the same approach as the Newtonian treatment. The local tidal tensor is evaluated in Boyer-Lindquist coordinates and then transformed into an orthonormal coordinate basis freely falling with the star. This ``local" coordinate system has spatial coordinates directed along the $\{ r, \theta, \phi \}$ axes. The only non-zero spatial elements of the tidal potential can be expressed in the form:
\begin{equation}
\phi_{, xx} = - \frac{1+k}{r^3},
 \phi_{, yy}= \frac{k}{r^3},
  \phi_{, zz} =\frac{1}{r^3},
\end{equation}
where 
\begin{equation}
k = \frac{\tilde{r}^2-4 a \tilde{r}^{\frac{1}{2}}+3a^2}{\tilde{r}^2-3\tilde{r}+2a\tilde{r}^{\frac{1}{2}}}.
\end{equation}
This satisfies Laplace's equation as it must. Note the remarkable prospects that $k\rightarrow2$ for $r \rightarrow r_{\rm{ISCO}}$ independent of $a$. Also $k \rightarrow 1$ as $r \rightarrow \infty$ as must also be true in the Newtonian limit.

Combining the stellar gravity and rotation with the tidal contribution, the Roche potential becomes:
\begin{eqnarray}
\Phi (x, y, z) &=& -m_\star (x^2 + y^2 + z^2)^{-\frac{1}{2}}  \nonumber \\
                       &&+ \frac{M}{r^3}  \frac{ ky^2 + z^2 - (k + 1) x^2}{2}   \nonumber \\
                       &&  -  \frac{M}{r^3}  \frac{k_\omega(x^2 + y^2)} {2}.
\end{eqnarray}
Here $k_\omega$ is the stellar rotational parameter in the inertial frame. 

Similarly to the Newtonian treatment, we can locate the first and second Lagrange points and the transverse extent of the Roche surface, and calculate the volume contained by this surface. When $k=2$, the Roche volume at the ISCO is calculated to be:
\begin{equation}
 \label{Vms} 
  V_{\rm {ISCO}} =  0.456  \times \left ( \frac{k_\omega}{4.09} +1 \right )^{-1} \frac{m_\star}{M} r^3.
\end{equation}
And more generally, an adequate approximation to the Roche volume is:
\begin{equation}
 \label{Rochevolumerel} 
  V(r, a, k_\omega) = V_{\rm N}(r, k_\omega) - \frac { V_{\rm N}(r, k_\omega)-V_{\rm ISCO}(r, k_\omega) } { \left( \frac{r}{r_{\rm ISCO}} \right)^{0.5}+ F(a, k_\omega)( \frac{r} {r_{\rm ISCO} }-1) }, \\
\end{equation}
where
\begin{eqnarray}
 F(a, k_\omega) &=& -23.3 + \frac{13.9}{k_\omega+ 2.8} \nonumber \\
                              && + \left (23.8 - \frac{14.8}{2.8 + k_\omega} \right)\times(1 - a)^{0.02}  \nonumber \\
                              &&  + \left(0.9 - \frac{0.4} {2.6 + k_\omega} \right) \times(1 - a)^{-0.16}  \nonumber. 
 \end{eqnarray}
 
\begin{figure}
 \centering
 \includegraphics[width=3in]{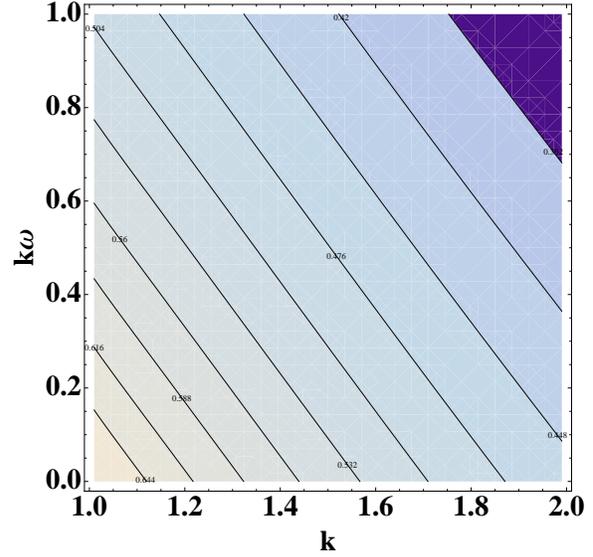}
 \caption{Contour plot of Roche volume as a function of $k$ and $k_\omega$ for $a=0.5$. $k=1$ is the Newtonian limit. $k=2$ when the star is at ISCO. $k_\omega = 0$ when the star is non rotating, and $k_\omega=1$ when the star is co-rotating.
 } 
 \label{contourV5}
\end{figure}

As an example, Fig. \ref{contourV5} is the contour plot of the relativistic Roche volume (in units of ${m_\star}/{M} r^3$) around a black hole with $a=0.5$ as a function of $k$ and $k_\omega$. This volume decreases from 0.683 to 0.456 from the Newtonian limit to the most extreme relativistic limit ($r=r_{\rm ISCO}$) for a non-rotating star. For a co-rotating star, the variation is from 0.503 ($k=1$) to 0.368 ($k=2$). These volumes at the four corners of the plot are independent of the spin.

\begin{figure}
 \centering
 \includegraphics[width=3in]{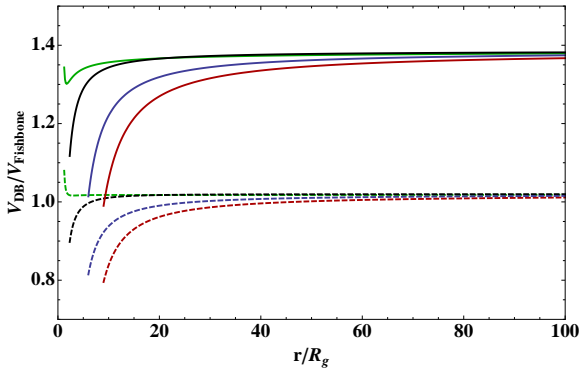}
 \caption{The ratio of our Roche volume to \citet{Fishbone73}'s Roche volume, as a function of $r / R_g$. The our four solid curves are for non-rotating stars orbiting a black hole with $a = -1$ (red), 0 (blue), 0.9 (black), and 0.998 (green). The lower four dashed ones are for co-rotating stars correspondingly.
 } 
 \label{RocheVRatio}
\end{figure}

Comparing with \citet{Fishbone73}'s Roche volume used by \citet{Hameury94}, our model is more accurate in the relativistic limit, including black hole's spin and the stellar rotation. We plot the ratio of two Roche volumes in Fig.\ref{RocheVRatio}

Using Eq. [\ref{AngularVelocity}] and Eq. [\ref{Rochevolumerel}], we obtain a relationship between the stellar mean density and the orbital radius when the star fills its Roche lobe:
\begin{eqnarray}
 \label{Rochedensity} 
  \frac {M}{\bar{\rho}_\star r^3} & = & - \frac {0.683  \times \left ( \frac{k_\omega}{2.78} +1 \right )^{-1}-0.456  \times \left ( \frac{k_\omega}{4.09} +1 \right )^{-1} } { \left( \frac{r}{r_{\rm ISCO}} \right)^{0.5}+ F(a, k_\omega)( \frac{r} {r_{\rm ISCO} }-1) }\nonumber \\
    && +0.683  \times \left ( \frac{k_\omega}{2.78} +1 \right )^{-1} .
\end{eqnarray}
Then we know how the stellar properties govern the orbital evolution during the Roche accretion phase.

\subsection{Roche evolution}
We studied how a star evolved adiabatically in \citet{Dai11a}, and calculated how the stellar mean density changes as mass is stripped for a representative set of stars. As the star approaches a SMBH in a circular, equatorial orbit, it might plunge into the SMBH directly, go through full Roche evolution, or start the Roche evolution then plunge into the hole. The detailed evolution depends on the mass and spin of the black hole, the stellar type, and the rotational parameter $k_\omega$ of the star. Using Eq. [\ref{Rochedensity}], we can check when the star will start Roche accretion and if so, how its orbit will change through the process.We summarize these inspiral/outspiral scenarios in Fig. \ref{RochePeriodMass}. 

Each plot shows how different stars behave as they spiral to within  $50 R_g$ of a SMBH with given mass and spin. General relativity plays an important role in this region. For example, in plot (f), the non-spinning SMBH has mass $10^7 M_\odot$. Brown dwarfs and white dwarfs will plunge into the hole. A Sun-like star, an upper main-sequence star with an initial mass $\sim 7.9 M_\odot$, and a lower main-sequence star with an initial mass $\sim 0.3 M_\odot$ will go through full Roche accretion. A red supergiant with an initial mass $\sim 11.85 M_\odot $ will start Roche accretion and will plunge into the hole.

The plots also show how the stellar orbital periods change in the Roche phase. The Period-Mass curves can be quite distinct with or without relativity or if the star is co-rotating with the black hole or not. In plot (c), a Sun-like co-rotating star barely crosses the ISCO of a $10^8 M_\odot$ SMBH with $a=0.9$ using Newtonian calculation so it can go through Roche evolution. However, using relativistic calculation this star will just plunge into the hole. And in plot (l), only with general relativity a co-rotating white dwarf with initial mass $0.8 M_\odot$ can start the Roche accretion phase near a maximally spinning M6 black hole. 

\begin{figure}
 \centering
 \includegraphics[width=3in]{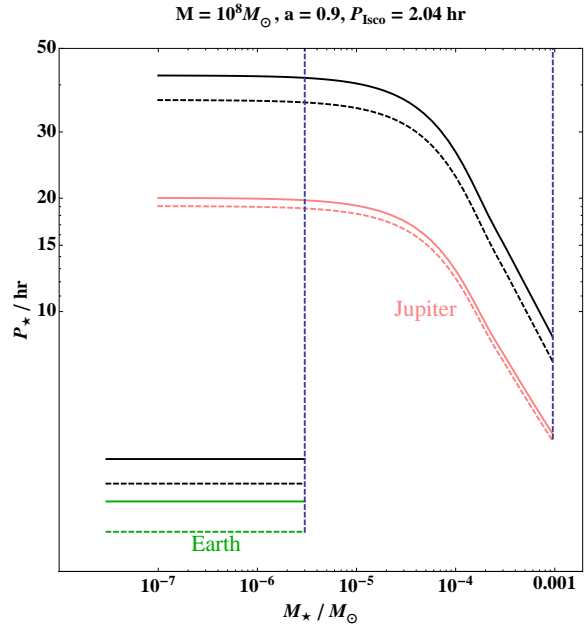}
 \caption{Roche evolutions of a perfect solid planet with the mass of Earth (green) and a perfect gas planet with the mass of Jupiter (pink) near a maximally spinning SMBH with $M = 10^8 M_\odot$. We plotted four curves for each planets for comparison: the relativistic calculation for a co-rotating star (colored solid curve), the relativistic calculation for a non-rotating star (colored dashed curve), the Newtonian calculation for a co-rotating star (corresponding black solid curve), and the Newtonian calculation for a non-rotating star (corresponding black dashed curve). The vertical blue dashed lines represent the phase of inspiral due to gravitational radiation. }
\label{Planet}
\end{figure}

\begin{figure}
 \centering
 \includegraphics[width=7in]{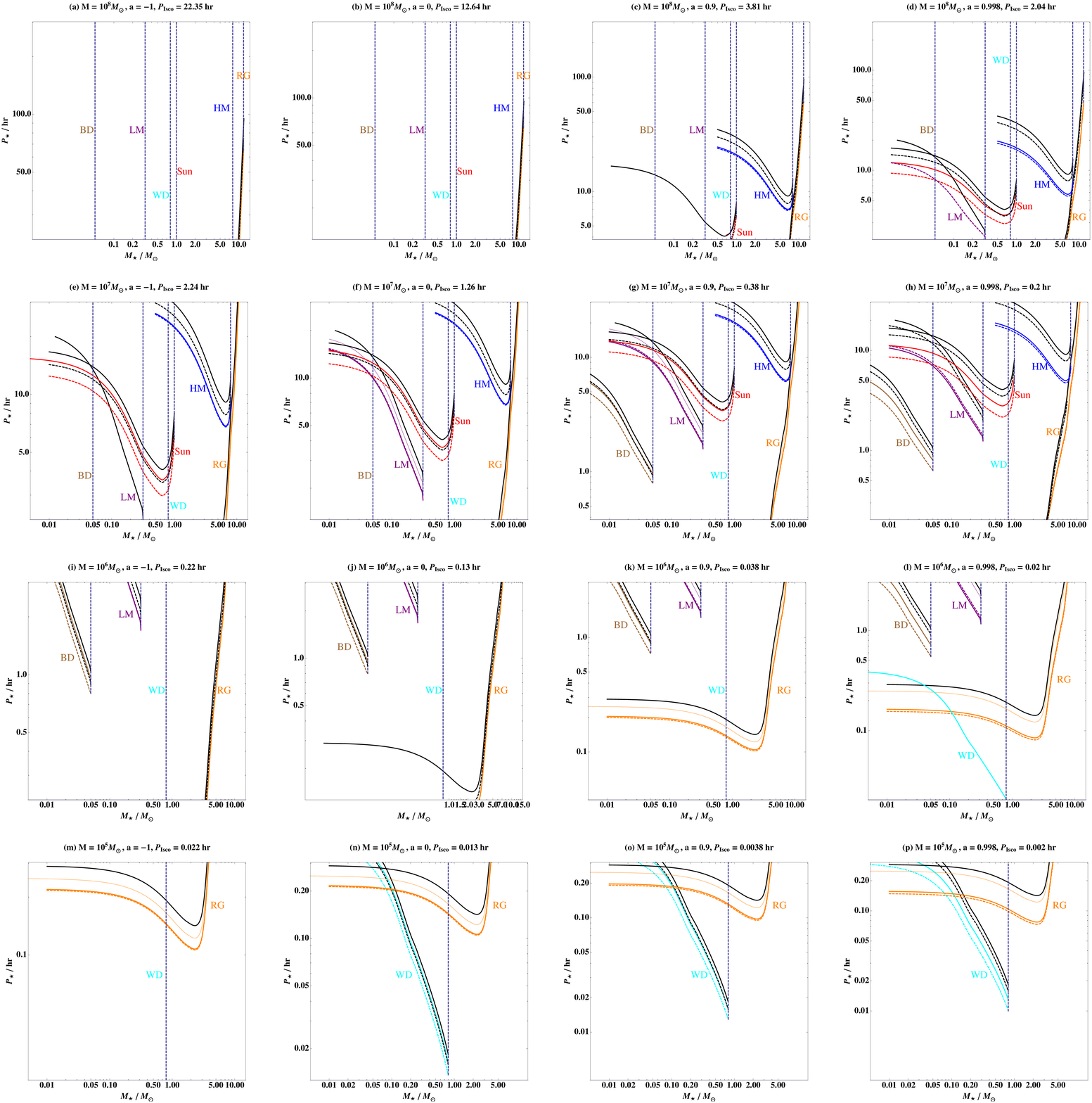}
 \caption{Roche evolutions of different stars close to SMBHs with $M = 10^8 M_\odot$, $10^7 M_\odot$, $10^6 M_\odot$, $10^5 M_\odot$, and $a = -1$, 0, 0.9 and 0.998. In each plot, the x-axis is the stellar mass ranging from $0.005 M_\odot$ to $15 M_\odot$, and the y-axis is the stellar orbital period from $P_{\text{ISCO}}$ to the orbital period at $50 R_g$ (close enough to the SMBH so these are the relativistic regions). We use different colors to denote different types of stars: a Sun-like star (red), a lower main sequence star with a mass $\sim 0.3 M_\odot$ (purple), an upper main sequence star with a mass $\sim 7.9 M_\odot$ (blue), a red super giant with a mass $\sim 11.85 M_\odot $(orange), a white dwarf with a mass $0.8 M_\odot$ (cyan), and a brown dwarf with a mass $0.05 M_\odot$ (brown). We plotted four curves for each star for comparison: the relativistic calculation for a co-rotating star (colored solid curve), the relativistic calculation for a non-rotating star (colored dashed curve), the Newtonian calculation for a co-rotating star (corresponding black solid curve), and the Newtonian calculation for a non-rotating star (corresponding black dashed curve). The vertical blue dashed lines represent the phase of inspiral due to gravitational radiation. }
 
\label{RochePeriodMass}
\end{figure}

We also plot the Period-Mass relationship for planets as in Fig. \ref{Planet}. A solid planet like Earth has the same period as mass decreases due to unchanging density. A gas giant like Jupiter evolves so that the period will increase. 

From the conservation of angular momentum, the stellar angular momentum $L$ satisfies:
\begin{equation}
 \label{Lrel} 
  \Delta L = \Delta L_{\rm GR} + \Delta L_{\rm ISCO} + \Delta L_{\rm wind} + \Delta L_{\rm rad}.
\end{equation}

Here $L$ is the angular momentum of the star as in Eq. [\ref{Lr}]. $\Delta  L_{\rm GR}$ is caused by the gravitational radiation torque $\tau_{\rm gr}$ calculated from out new model (Eq .[\ref{torquer}]). $\Delta L_{\rm ISCO}$ represents the loss of angular momentum of materials falling into the ISCO. $\Delta L_{\rm wind}$ is the angular momentum of materials lost from an accretion disc. In this inner region of an AGN accretion discs, there are mainly two kinds of wind: radiation pressure and magnetic driven. The radiation pressure induced wind carries its own specific angular momentum as on the disc. The magnetic induced wind carries away very little mass but very large angular momentum which can be 10 times as large as its specific angular momentum on the disc. The last term is the disc radiation torque due to beaming effect, which is relatively small compared with the others.

\onecolumn

\twocolumn

In order to understand the details of $\Delta L_{\rm {wind}}$, a three-dimensional dynamical simulation will need to be employed. In this paper, for simplicity, we consider only the first two $\Delta L$'s, the mass accretion rate onto the hole will be the same as the stellar mass deposition rate. We then have:
\begin{equation}
 \label{Lrel2} 
  \frac{dL}{dt} + \tau_{\rm{gr}} +  \dot{m}_\star j_{\rm ms} = 0.
\end{equation}
$j_{\rm ms}$ is the specific angular momentum at the ISCO, which equals 0 in the Newtonian limit.

For the same dwarf star with mass $\sim 0.3 M_\odot$, we can calculate how its $P$ and $\dot m$ change in the relativistic limit in comparison to the Newtonian treatment, as in Fig .\ref{PMlm} and Fig. \ref{Mdotlm}. With general relativity, the star does not recede from the hole as far as in the Newtonian limit, and the mass accretion rate has dropped to about half the Newtonian rate. 

We can apply the same method as in Section 2.3 to check if the stellar volume will exceed the Roche volume in the evolution. Contradictory to \citet{Hameury94}'s result, the Roche evolution of a dwarf star is actually stable in the relativistic limit on dynamical time scales.

  \section{Constraining spins from the signals}
  
  Returning to the observation of RE J1034+396, the $\sim1$hr periodic X-ray signal can be associated with a red giant or lower main-sequence star or brown dwarf orbiting and accreting onto a $\sim10^6 M_\odot$ black hole or a $ \sim10^7 M_\odot$ hole (the latter with a high spin). 16 cycles of signals are insufficient to be confident of a stable period. However, if more signals can be observed, information on spin could be extracted. In principle, the closer the star is to the hole, or the larger the spin of the hole, the stronger the emission would be. This could introduce an important selection effect as rapidly spinning holes would be more likely to be observed.
  
\begin{figure} 
\centering
 \includegraphics[width=3in]{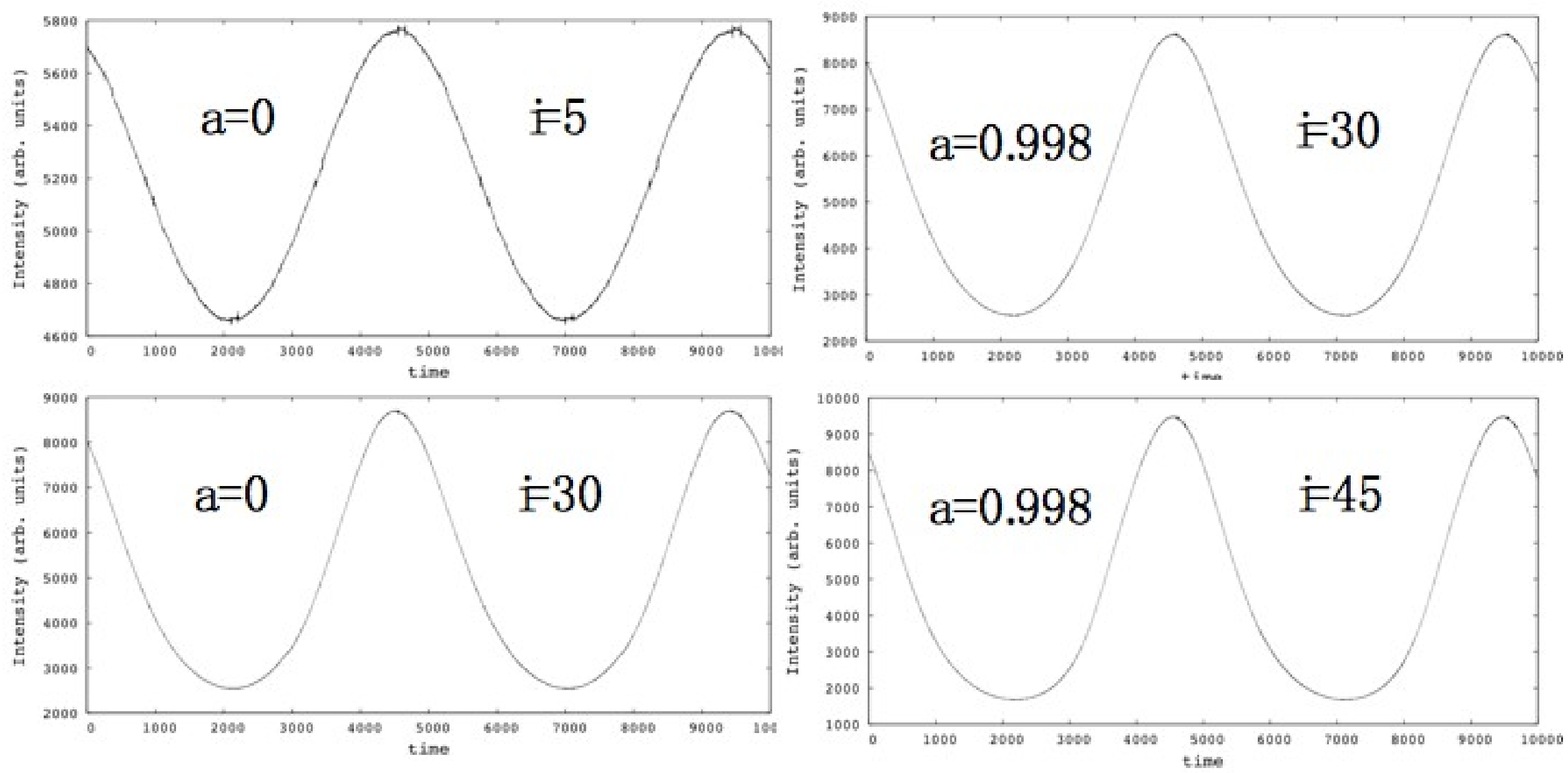}
 \caption{Simulated X-ray period emission from a hot spot orbiting a black hole is shown for various black hole spin parameter $a$ and observer inclination angle $i$. In each figure the hot spot orbits the hole in equatorial circular orbits with the same radius. The units are arbitrary but consistent.}
\label{Hotspotcurve}
\end{figure}

  Fig. \ref{Hotspotcurve} shows simulated periodic signals from a hot spot orbiting around the hole circularly using our numerical package. The modulation of intensity comes from beaming effect - as we discussed in \citet{Dai10}, the bolometric intensity varies as the fourth power of the Doppler-gravitational redshift factor $g=E/E_{\rmn{0}}$ \citep{Cunningham75}. Here $E$ is the energy of the photon in the observer's frame and $E_{\rmn{0}}$ in the hot spot's inertial rest frame.The shape of the curves and the relative intensity are different for different spins and observer inclination angles. 
  
  Also a major difference between producing a hot spot on the disc using the Roche accretion model and other QPO models (e.g. \citet{Bob01}) is that the periodicity of the modulation will follow that of the stellar orbit although there is likely to be phase noise. Since the stellar orbit evolves on an adiabatic time scale, the periodicity of the hot spot will also change very slowly but in a consistent way.  For example, an accreting $\sim 0.3 M_\odot$ dwarf star's orbital periodicity can change by as much as $\sim 0.1$s in one year, in which time we could observe $\sim 8000$ cycles of modulation.  A $P-\dot P$ analysis similar to the double pulsar's could in principle be done to extract information of the stellar type and black hole mass and spin.

  \section{Discussion and future research directions}
  
   Observation of quasi-periodic X-ray emissions from stars orbiting black holes in Active Galactic Nuclei (AGN) is a potential probe of general relativity. There is an intriguing possibility that the X-ray modulation of RE J1034+396 does come from an orbiting star filling its Roche lobe and being swallowed b the black hole, in a similar fashion to the "black widow" pulsar systems (e.g. \citet{Thorsett94}) like PSR1957+20 (e.g. \citet{Kulkarni88, Kulkarni92, Arons93}). It is reasonable to search the existing AGN X-ray data base for periodic signals with relevant periods within range $\sim1-20$hr.

   Although we are mostly interested in this paper in observable tidally disrupted or accreting stars, it is quite reasonable to suppose that most dormant AGN black holes are orbited by a stable system of stars with a common orbital plane in much the same way as the sun is orbited by the planets. Sgr A$^\star$, which already has over 20 closely orbiting stars, might eventually evolve to such a configuration. Other orbiting stars can have also dynamical interactions with the accreting star and make its evolution more complicated. 
            
   The probability of observing such phenomena relates to the stellar distribution in accretion discs and the mechanisms of bringing stars close to ISCO. When material accrete onto the black holes, stars can be borne in the accretion disc, which could go away eventually leaving only the stars. Such stars would orbit the hole in bound orbits in the equatorial plane with low eccentricity. On the other hand, if an accretion disc still exists when the star gets close to ISCO, the star will have interaction with the disc as well, or might be traveling in the disc already producing a gap and perturbation patterns on the disc. 
   
   A series of other interesting future projects can also be done on this model. If the star is on a slightly eccentric orbit or inclined orbit, the accretion rate could be larger than the case of a circular orbit, but perturbations will be added to the sinusoidal behavior of the flux. A main-sequence star on its way to become a red-giant can also increase the mass accretion rate. A jet can also be formed due to the Roche accretion - the radio component will also be very useful in providing information on the black hole parameters. Between a total tidal disruption and a steady mass-transfer, intermediate phases can also exist and we think much can be done on that. In order to understand the stellar evolution and radiation mechanisms better, we also need to do a three-dimensional general relativistic dynamical simulation. The time it would take to run such evolutions on existing numerical codes would be too long. New numerical methods need to be designed. 
   
   However, the most exciting and immediate prospect is to open up a new line of investigation using existing X-ray databases, to seek evidence for stars in bound orbits around massive black holes. Discovering such systems would elucidate the evolution of AGN and also open the possibility of a new measurement of black hole spin and ultimately of concluding that the Kerr metric correctly describes spinning black holes.

\section*{Acknowledgments}

This work was supported by the U.S. Department of Energy contract to SLAC no. DE-AC02-76SF00515. We would like to give special thanks to our collaborator P. Eggleton for providing the stellar models, R. Wagoner for valuable comments, and D. Chakrabarty, J. Faulkner, L. Gou, J. Grindlay, S. Hughes, A. Loeb, H. Marshall, J. McClintock, M. Nowak, S. Phinney, R. Remillard, P. Schechter, R. Simcoe, A. Tchekhovskoy, and N. Weinberg for useful discussions.

\end{document}